\documentclass[twocolumn,showpacs,showkeys,preprintnumbers,
amsmath,amssymb,floatfix,aps,pre]{revtex4}

\usepackage{graphicx}
\usepackage{amsmath}
\usepackage{multirow}
\usepackage{times}
\usepackage{amssymb}
\usepackage{dcolumn}%
\usepackage{bm}%
\usepackage[]{hyperref}%
\usepackage{upgreek}
\usepackage{wasysym}
\usepackage{xcolor}

\hypersetup
{
	colorlinks=true,
	linkcolor=blue,
	citecolor=blue
}

\begin{document}

\title[]{Weak links between fast mobility and local structure in molecular and atomic  liquids}
\author{S.\@ Bernini}
\affiliation{Dipartimento di Fisica ``Enrico Fermi'', 
Universit\`a di Pisa, Largo B.\@Pontecorvo 3, I-56127 Pisa, Italy}
\author{F.\@ Puosi}
\affiliation{Laboratoire de Physique de l'\'Ecole Normale Sup\'erieure de Lyon,  UMR CNRS 5672, 46 all\'ee d'Italie, 69007 Lyon, France}
\author{D.\@ Leporini}
\email{dino.leporini@df.unipi.it}
\affiliation{Dipartimento di Fisica ``Enrico Fermi'', 
Universit\`a di Pisa, Largo B.\@Pontecorvo 3, I-56127 Pisa, Italy}
\affiliation{IPCF-CNR, UOS Pisa, Italy}

\date{\today}

\begin{abstract}
\noindent
We investigate by Molecular-Dynamics simulations the fast mobility -  the 
rattling amplitude of the particles temporarily trapped by the cage of the 
neighbors - in mildly supercooled states of dense molecular 
(linear trimers) and atomic (binary mixtures) liquids. {The mixture particles 
interact by the Lennard-Jones potential. The non-bonded particles of the 
molecular system are coupled by the more general Mie potential with variable 
repulsive and attractive exponents in a range which is characteristic of small 
$n$-alkanes and $n$-alcohols}. 
Possible links between the fast mobility and the geometry of the cage  (size 
and shape) are searched.  The correlations on a per-particle basis are rather 
weak. Instead, if  one groups either the particles in fast-mobility subsets or 
the cages in geometric subsets, the increase of the fast mobility with both the 
 size and the asphericity of the cage is revealed. The observed correlations 
are weak and differ in states with equal relaxation time. 
Local forces between a tagged particle and the first-neighbour 
shell do not correlate with the fast mobility in the molecular liquid. It is 
concluded that the cage geometry alone is unable to provide a microscopic 
interpretation of the known, universal link between the fast mobility and the 
slow structural relaxation. We suggest that the particle fast dynamics is 
affected by regions beyond the first neighbours, thus supporting the presence 
of collective, extended fast modes.
\end{abstract}

\maketitle

\section{Introduction}

Understanding the extraordinary viscous slowing down that characterizes  the 
glass formation is a major scientific
challenge \cite{DebeStilli2001,BerthierBiroliRMP11,EdigerHarrowell12}.  From a 
microscopic point of view, when a liquid approaches the glass transition, the 
particles are subjected to prolonged trapping by their surroundings and the 
structural relaxation time $\tau_\alpha$, i.e the average escape time from the 
cage of the first neighbors, increases from a few picoseconds up to thousands 
of seconds. During the trapping period the rattling motion inside the cage 
occurs on picosecond time scales with mean square amplitude $\langle u^2 
\rangle$. Henceforth $\langle u^2 \rangle$ will be referred to as "fast 
mobility". 
The temporary trapping and subsequent escape mechanisms lead to large  
fluctuations around the averaged dynamical behavior with strong heterogeneous 
dynamics \cite{BerthierBiroliRMP11} and 
{non-exponential relaxation \cite{MonthBouch96}}.
Pioneering \cite{TobolskyEtAl43,RahmanCageCentroidJCP66} investigations  and 
recent theoretical 
\cite{Angell95,HallWoly87,Dyre06,MarAngell01,Ngai04,Ngai00,
DouglasCiceroneSoftMatter12,WyartPNAS2013,
SchweizerElastic1JCP14,SchweizerElastic2JCP14,Elastico2} 
 studies addressed the  rattling process in the cage to understand the  
structural relaxation - the escape process -  gaining support from numerical 
\cite{Angell68,Nemilov68,Angell95B,StarrEtAl02,BordatNgai04,Harrowell06,
Harrowell_NP08,HarrowellJCP09,DouglasEtAlPNAS2009,XiaWolynes00,DudowiczEtAl08,
DouglasCiceroneSoftMatter12,OurNatPhys,lepoJCP09,Puosi11,SpecialIssueJCP13,
UnivSoftMatter11,CommentSoftMatter13,DouglasStarrPNAS2015,OttochianLepoJNCS11,
UnivPhilMag11,Puosi12SE,Puosi12,PuosiLepoJCPCor12,PuosiLepoJCPCor12_Erratum,
Elastico2} and experimental works on glassforming liquids 
\cite{BuchZorn92,AndreozziEtAl98,Cicerone11,DouglasCiceroneSoftMatter12,
SokolovNovikovPRL13}.  

Renewed interest about the fast mobility was raised by extensive molecular-dynamics (MD) simulations evidencing the universal correlation between the structural relaxation time $\tau_\alpha$ and $\langle u^2\rangle$  were reported in polymeric systems \cite{OurNatPhys,lepoJCP09,Puosi11}, binary atomic mixtures \cite{lepoJCP09,SpecialIssueJCP13}, colloidal gels \cite{UnivSoftMatter11} and antiplasticized polymers \cite{DouglasCiceroneSoftMatter12,DouglasStarrPNAS2015}
and compared with the experimental data concerning several glassformers in   a 
wide fragility range ($20 \le m \le 191$) 
\cite{OurNatPhys,UnivPhilMag11,OttochianLepoJNCS11,SpecialIssueJCP13,
CommentSoftMatter13}.
One major finding was that states{of a given physical system}, say X and Y,  
with equal fast mobility  $\langle u^2\rangle$ have equal relaxation times 
$\tau_\alpha$ too:
\begin{equation}
\langle u^2 \rangle^{(X)}  =  \langle u^2 \rangle^{(Y)}  \iff \tau_{\alpha }^{(X)} = \tau_{\alpha }^{(Y)}
\label{vanhovescaling2}
\end{equation}
{The generalisation of Eq.\ref{vanhovescaling2} to deal with states of  
different physical systems is discussed elsewhere 
\cite{OurNatPhys,lepoJCP09,Puosi11,SpecialIssueJCP13} and exemplified in 
Sec.\ref{dynamic}}.
{After proper account of  the molecular-weight dependence} 
Eq.\ref{vanhovescaling2}  extends to longer time scales in polymers where 
states with equal fast mobility exhibit also equal chain reorientation rate 
\cite{lepoJCP09,Puosi11} and diffusivity \cite{Puosi11,Puosi12SE}. Correlation 
between diffusion and fast mobility was observed in atomic mixtures too 
\cite{SpecialIssueJCP13}. 
A more general relation leading to Eq.\ref{vanhovescaling2}  is provided by  
the particle displacement distribution, i.e. the incoherent, or self part, of 
the van Hove function $G_{s}(r,t)$  
\cite{Egelstaff:1992fk,HansenMcDonaldIIIEd}. The interpretation of $G_{s}(r,t)$ 
is direct. The product $G_{s}(r,t) \cdot 4\pi r^{2} \, dr$ is the probability 
that the particle is at a distance between $r$ and $r+dr$ from  the initial 
position after a time $t$. In terms of  $G_{s}({ r},t)$ the relation between 
fast and slow dynamics is expressed by stating that, if two  states{of a given 
physical system} are characterized by the {\it same} displacement distribution 
$G_{s}( r,t^*)$ at the trapping time $t^*$, they also exhibit the {\it same} 
distribution at long times, e.g. at $\tau_\alpha$ 
\cite{Puosi11,SpecialIssueJCP13}:
 \begin{equation}
G_{s}^{(X)}({ r},t^*) = G_{s}^{(Y)}({ r},t^*)   \iff G_{s}^{(X)}({ 
r},\tau_\alpha)  = G_{s}^{(Y)}({r},\tau_\alpha)
\label{vanhovescaling}
\end{equation}
The precise definition of the trapping time $t^*${(not to be confused with  the 
peak position of the non-Gaussian parameter)} will be given in 
Sec.\ref{dynamic} where it  will be also shown that Eq.\ref{vanhovescaling} 
implies Eq.\ref{vanhovescaling2}. 
Actually, the coincidence of the two incoherent van Hove functions extends  
across all the time interval between $t^*$ and at least $\tau_\alpha$ 
\cite{Puosi11,SpecialIssueJCP13}.
Eq.\ref{vanhovescaling} holds even in the presence of very strong dynamical  
heterogeneity  where both diffusive and jump-like dynamics are observed 
\cite{Puosi11}. From this respect, it offers an interpretation of the scaling 
of the breakdown of the Stokes-Einstein law in terms of the fast mobility 
\cite{Puosi12SE} and is consistent with previous conclusions that the long-time 
dynamical heterogeneity is predicted by the fast heterogeneities 
\cite{Harrowell06,note3}.

Several approaches  suggest that structural aspects matter in the dynamical  
behavior of glassforming systems. This includes the Adam-Gibbs derivation of 
the structural relaxation \cite{AdamGibbs65,DudowiczEtAl08} - built on the 
thermodynamic notion of the configurational entropy \cite{GibbsDiMarzio58} -, 
the mode- coupling theory \cite{GotzeBook} and extensions 
\cite{SchweizerAnnRev10}, the random first-order transition theory 
\cite{WolynesRFOT07}, the frustration-based approach \cite{TarjusJPCM05}, as 
well as the so-called  elastic models  
\cite{Dyre06,Puosi12,DyreWangGpJCP12,Elastico2,WyartPNAS2013,
DouglasStarrPNAS2015,SchweizerElastic1JCP14,SchweizerElastic2JCP14}.  It was 
concluded that the proper inclusion of many-body static correlations in 
theories of the glass transition appears crucial for the description of the 
dynamics of fragile glass formers \cite{coslovichPRE11}.
The search of a link between structural ordering and slow dynamics motivated  
several studies in liquids 
\cite{NapolitanoNatCom12,EdigerDePabloNatMat13,RoyallPRL12,BarbieriGoriniPRE04,
ReichmannCoslovichLocalOrderPRL14}, colloids 
\cite{StarrWeitz05,TanakaNatMater08,TanakaNatCom12} and polymeric systems 
\cite{StarrWeitz05,DePabloJCP05,GlotzerPRE07,LasoJCP09,BaschnagelEPJE11,
MakotoMM11,LariniCrystJPCM05}.  { We also note that the elastic models of the 
glass transition conclude that the relaxation is related to the elastic shear 
modulus, a quantity which is set by the  arrangement of the particles in 
mechanical equilibrium and their mutual interactions \cite{Dyre06,Puosi12}.}

{The present work is carried out in the spirit of the cell theory of the  
liquid state \cite{Egelstaff:1992fk}. Each cell is identified by the Voronoi 
polyhedron (VP) which, as originally suggested by Rahman 
\cite{RahmanCageCentroidJCP66} and later studies \cite{StarrEtAl02}, provide 
useful information about the local molecular environment, e.g. to get to the 
equation of state of the hard sphere liquid \cite{SastryEtAl98}. 
For a given configuration, the VP surrounding a particle encloses all the  
points which are closer to it than to any other one.}
Following a preliminary report \cite{VoronoiBarcellonaJNCS14}, we investigate  
by thorough Molecular-Dynamics (MD) simulations the particle fast mobility and 
relaxation of linear molecules and binary atomic mixtures. The purpose is the 
assessment of the correlations of the fast mobility $\langle u^2 \rangle$ with 
the size and the shape of the cage formed by the nearest neighbors of the 
trapped particle. The local geometry is characterized by the volume and the 
asphericity of the associated VP. {Alternatives are reported 
\cite{DouglasCiceroneSoftMatter12}}.  Asphericity has been considered in 
polymers
\cite{SegaJCP,StarrEtAl02,DouglasCiceroneSoftMatter12,Rigbyt,
VoronoiBarcellonaJNCS14},  water 
\cite{StarrEtAl02,H2SJCP,ShihAsphericityWater,JedloJCP,Wikfeldt10,Stirnemann12}
, silica \cite{StarrEtAl02}, small molecules \cite{H2SJCP}, packed beds of 
particles \cite{LiaoAsphericityPowderTech02}, hard  \cite{Krekelberg06} and 
soft \cite{MontoroAbascal93} spheres and its better ability to discriminate 
different local structures with respect to other features of the Voronoi 
polyhedron has been noted \cite{MontoroAbascal93,LiaoAsphericityPowderTech02}. 
 
{The motivation for searching for a connection between cage geometry and  fast 
mobility is that it is intuitively expected. Nonetheless, we}  find that the 
correlations are rather weak on a per-particle basis. 
More insight is provided 
by grouping the ensemble of particles in (nearly) iso-volume, iso-asphericity  
and iso-mobility subsets which reveals the increase of the fast mobility in 
cages with increasing size and asphericity. However, the correlations are weak 
and differ in states with equal relaxation time, thus preventing the 
microscopic interpretation of Eq.\ref{vanhovescaling2}, Eq.\ref{vanhovescaling} 
in terms of the cage geometry alone. 
It is found that the local forces between a tagged particle  
and the first-neighbour shell do not correlate with the fast mobility in the 
molecular liquid. 
We offer arguments suggesting that the single-particle fast mobility is  
affected by regions extending beyond the nearest neighbours in agreement with 
other studies  
\cite{Dyre06,Puosi12,DyreWangGpJCP12,Elastico2,WyartPNAS2013,
DouglasStarrPNAS2015,SchweizerElastic1JCP14,SchweizerElastic2JCP14,
WyartJStatMech07,Harrowell06,Harrowell_NP08,HarrowellJCP09,LemaitrePRL14,
BerthierJackPRE07,VoronoiBarcellonaJNCS14}. 

The paper is organized as follows. In Sec.\;\ref{numerical} the MD  algorithms 
are outlined, and the atomic and the molecular models are detailed. The results 
are presented and discussed in Sec.\;\ref{resultsdiscussion}. In particular, in
Sec.\;\ref{dynamic} the fast rattling motion and the long-time relaxation  are 
characterised, whereas Sec.\;\ref{correlations}{ expounds for the first time on 
the search of cross correlations between the cage geometry and the fast 
dynamics}. Finally, the conclusions are summarized
in Sec. \ref{conclusions}.

\section{Methods}
\label{numerical}

\subsection{Molecular liquid}
\label{polymer}

A coarse-grained model of fully-flexible linear chains of trimers is used. 
Nonbonded monomers at a distance $r$ interact via
the truncated parametric{Mie} potential \cite{MiePot}:

\begin{equation}
\label{parametric}
U_{p,q}(r)=\frac{\varepsilon}{q-p} \left[ p \left( \frac{\sigma^*}{r}\right)^q
-q\left(\frac{\sigma^*}{r}\right)^p \right] + U_{cut} 
\end{equation}

Changing the $p$ and $q$ parameters does not change the position $r = \sigma^*  
= \sqrt[6]{2}\,\sigma$ and the depth of the potential minimum
$\varepsilon${but it affects the fragility of the system \cite{BordatNgai04}}.  
The constant $U_{cut}$ is chosen to ensure $U_{p,q}(r) = 0 $ at $r \geq 2.5 \, 
\sigma$. Note that
$U_{p,q}(r)=U_{q,p}(r)$. Conventionally $p<q$ and the choice $p=6,q=12$ yields  
the familiar Lennard-Jones (LJ) potential.{The Mie potential has been 
considered to model microscopic interactions in solids including ionic ones 
($p=1$, Born-Mie potential) \cite{StaceyMie}, metallic glasses 
\cite{WangEtAlMieMetGlas03} and polar molecules ($p \simeq 3$), atomic  
liquids\cite{LekkerkerkerMiePotPhysA99}, as well as simple and chain molecular 
\cite{MansooriMiePot80,JacksonMiePolymJCP97,JacksonMiePolymerJCP13} liquids.  
We considered the following set of $(p,q)$ pairs: 
$(7,15),(8,12),(7,11),(6,12),(6,8)$. 
The set, together with the choice of simulating liquids of short trimeric  
chains, is expected to be characteristic of simple non polar and polar 
molecular systems.  In fact,  the optimised parameters $(p,q,m_S)$ of the 
equation of state for chain molecules formed from $m_S$ tangentially-bonded  
Mie monomers are \cite{JacksonMiePolymerJCP13}: $(6,18.9,3)$ for n-decane, 
$(6,15.85,1.96)$ for n-pentane, $(6,11.7,2.44)$ for n-butan-1-ol and 
$(6,7.61,1.96)$ for ethanol. Notice that  the present work neglects the 
additional short-ranged square-well potential which is used for the alcohols  
in ref.\cite{JacksonMiePolymerJCP13}. }

\begin{figure}[t]
\begin{center}
\includegraphics[width=0.71\linewidth]{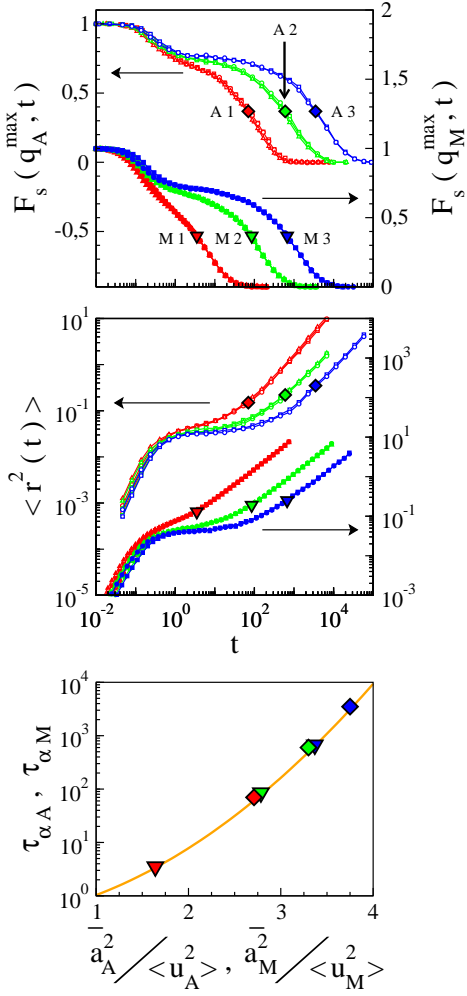} \\
\end{center}
\caption{Top: ISF of selected iso-relaxation time states of the atomic  ($A1$, 
$A2$ $A3$, left axis) and molecular ($M1$, $M2$, $M3$, right axis) liquids. The 
structural relaxation times of the atomic, $\tau_{\alpha A}$, and molecular, 
$\tau_{\alpha M}$, liquids are marked as diamonds and triangles, respectively.  
Middle: corresponding 
MSD. 
Note that the MSD 
 of each set of iso-relaxation time states coincide for $t\gtrsim 0.5$, i.e.  
their fast mobility is equal as stated by Eq.\ref{vanhovescaling2}. Bottom:
scaling of the relaxation time and the fast mobility. The continuous line is  
the master curve given in Eq.3 of ref. \cite{SpecialIssueJCP13} 
($\overline{a^2_A}=0.1124$ and $\overline{a^2_M} \equiv \overline{a^2_P} 
=0.1243$ \cite{SpecialIssueJCP13}). For the atomic liquid the iso-relaxation 
time states $(\rho,T)$ are: $A1$: $\color{red}
\Square$
(1.125,0.388); $\color{red} \Circle$ (1.204,0.55); $\color{red} \triangle$  
(1.296,0.804). $A2$: $\color{green} \Square$
(1.125,0.338); $\color{green} \Circle$ (1.204,0.48); $\color{green} \triangle$  
(1.296,0.702). $A3$: $\color{blue} \Square$
(1.125,0.317); $\color{blue} \Circle$ (1.204,0.45). For the molecular liquid 
the  iso-relaxation time states $(\rho,p,q,T)$ are: $M1$: $\color{red}
\blacksquare$ (1.015,7,15,1); $\color{red} \CIRCLE$ (1.070,6,8,0.55); 
$\color{red}  \blacktriangle$ (1.090,8,12,1.35). $M2$:
$\color{green} \blacksquare$ (1.018,7,15,0.7); $\color{green} \CIRCLE$ 
(1.070,6,8,0.39);  $\color{green} \blacktriangle$
(1.1,7,11,0.79). $M3$: $\color{blue} \blacksquare$ (0.984,6,12,0.33); 
$\color{blue}  \CIRCLE$ (1.086,6,12,0.63).}
 \label{ISF-MSD}
\end{figure}

\begin{figure}[t]
\begin{center}
\includegraphics[width=0.9\linewidth]{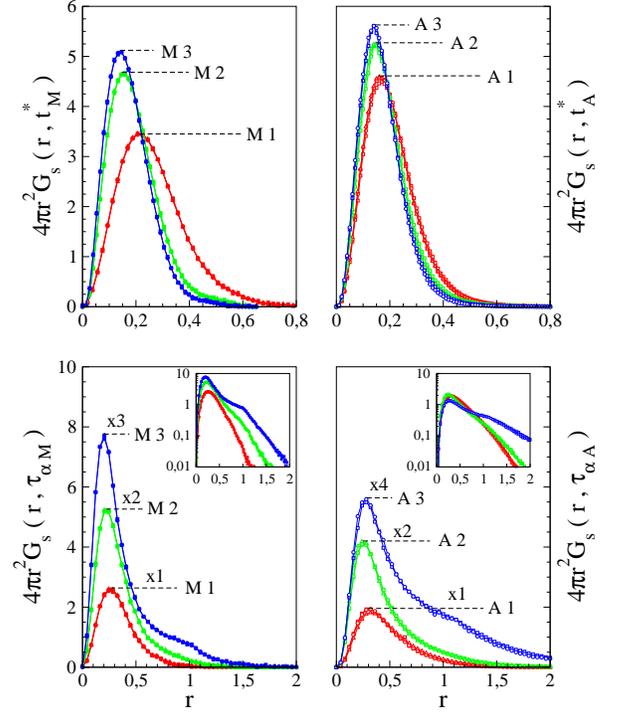} \\
\end{center}
\caption{Incoherent van Hove function of the iso-relaxation time states  
evaluated at the trapping (top) and the structural relaxation (bottom) times of 
the molecular (left) and the atomic  (right) liquids. The figure evidences the 
strong version of the scaling between the fast trapping regime and the slow 
relaxation one (Eq.\ref{vanhovescaling}).  Note that $G_s(r,\tau_{\alpha})$ of 
the sets of states M3 and A3 exhibits a bump at $r\sim 1$, the particle 
diameter, due to the jump dynamics characteristic of slowed-down states. The 
insets show that the coincidence of $G_s(r,\tau_{\alpha})$ extends to regions 
where the exponential tails  related to the dynamic heterogeneity are apparent 
\cite{KobPRL07}.}
 \label{vanHove}
\end{figure}

The bonded monomers
interact by a potential which is the sum of the LJ potential and the FENE  
(finitely extended nonlinear elastic) potential
\cite{sim}:
\begin{equation}
 U_{FENE}(r)=-\frac{1}{2}kR_0^2\ln\left(1-\frac{r^2}{R_0^2}\right)
\end{equation}
where $k$ measures the magnitude of the interaction and $R_0$ is the maximum  
elongation distance. The parameters $k$ and $R_0$
have been set  to $30 \, \varepsilon  / \sigma^2 $ and $ 1.5\,\sigma $  
respectively \cite{GrestPRA33}. The resulting bond length
is $b=0.97\sigma$ within a few percent. All quantities are in reduced units:  
length in units of $\sigma$, temperature in units of
$\varepsilon/k_B$ and time $\tau_{MD}$ in units of $\sigma \sqrt{\mu /  
\varepsilon}$ where $\mu$ is the monomer mass. We set $\mu
= k_B = 1$. 

We study systems of $N_M=2001$ monomers at different density $\rho$,  
temperature $T$ and $p,q$
parameters. Each state is labeled by the multiplet $(\rho,p,q,T)$. For  clarity 
reasons, the details about the states are given
in the caption of Fig.\;\ref{ISF-MSD}.

$NVT$ ensemble (constant number of particles, volume and temperature)  has been 
used
for equilibration runs, while $NVE$ ensemble (constant number of particles,  
volume and energy) has been used for production runs for a given
state point. $NVT$ ensemble is studied by the extended system method introduced 
 by Andersen \cite{Andersen80} and Nos\'e \cite{NTVnose}. The
numerical integration of the augmented Hamiltonian is performed through the 
multiple  time steps algorithm, reversible Reference System
Propagator Algorithm (r-RESPA) \cite{respa}. 

It is interesting to map the reduced MD units to real physical units. The 
procedure  involves
the comparison of the experiment with simulations and provide the basic length 
$\sigma$,  temperature $\varepsilon/k_B$ and time $\tau_{MD}$͒
units \cite{sim,KremerGrestJCP90,Kroger04,PaulSmith04,SommerLuoCompPhysComm09}. 
For example for  polyethylene and polystyrene it was found
$\sigma=5.3$ \AA, $\varepsilon/k_B = 443$ K ,$\tau_{MD} = 1.8$͒ ps and 
$\sigma=9.7$ \AA,  $\varepsilon/k_B = 490$ K ,$\tau_{MD} =
9$ ps respectively \cite{Kroger04}.

\subsection{Binary atomic mixture}
\label{mixture}

An 80:20 binary mixture of $N_{BM} = 4000$ particles is considered 
\cite{KobAndersenPRL}.  The two species $A,B$ interact via the potential:

\begin{equation}
U_{\alpha,\beta}(r)=\varepsilon_{\alpha,\beta} \left[\left( \frac{\sigma^*_{\alpha,\beta}}{r}\right)^{12}
-2\left(\frac{\sigma^*_{\alpha,\beta}}{r}\right)^6 \right] + U_{cut} 
\end{equation}

which is similar to Eq.\;\ref{parametric} with $p=6,q=12$ except that the 
height and  the minimum of the potential now depend
on the interacting species, being $\alpha,\beta \in \{A,B \}$ with 
$\sigma_{AA}=1.0$,  $\sigma_{AB}=0.8$, $\sigma_{BB}=0.88$,
$\varepsilon_{AA}=1.0$, $\varepsilon_{AB}=1.5$, $\varepsilon_{BB}=0.5$. We set 
the  masses $m_A=m_B=1$. Using argon units
for the A particles $\sigma_{AA}=3.405$ \AA, $\varepsilon_{AA}/k_B = 119.8 K$,  
$m_A=6.6337\cdot10^{-26}$ Kg, the time unit is
$\tau'_{MD}=(\sigma^2_{AA}m_{AA}/\varepsilon_{AA})^{1/2}=2.2$ ps 
\cite{AshwinSastry03}.  The system is equilibrated in the $NTV$
ensemble and the production runs are carried out in the $NVE$ ensemble. $NTV$ 
runs use a  standard and Nos\'e method
\cite{NTVnose}. The ``velocity verlet'' integration algorithm is used both in 
the $NVE$  and $NVT$ ensembles. We investigate only
the A particles of states with different number density $\rho$ and temperature 
$T$. Each  state is labeled by the
multiplet $(\rho,T)$. For clarity reasons, the details about the states are 
given in the  caption of Fig.\;\ref{ISF-MSD}.

\begin{figure}[t]
\begin{center}
\includegraphics[width=0.8\linewidth]{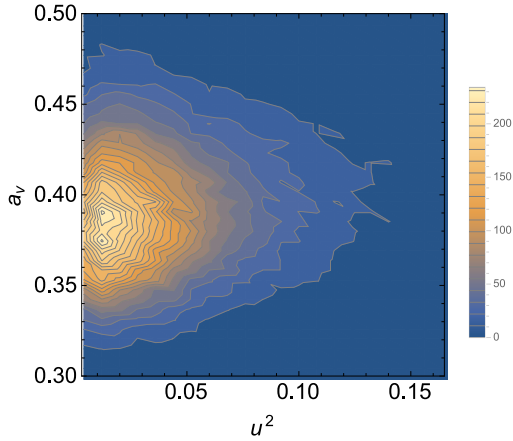} \\
\end{center}
\caption{Joint distribution of the VP asphericity $a_v$ and the fast mobility  
$u^2$,  $P(a_v,u^2)$, of the molecular-liquid state $\color{green} \CIRCLE$ 
(see caption of
Fig.\;\ref{ISF-MSD}). The two small peaks at $u^2 \simeq  0.012$ with $a_v   
\simeq 0.375$ and $a_v  \simeq 0.39$ are the peaks of the distributions of the 
central (higher asphericity) and the end (lower asphericity) monomers of each 
chain. Note that the peaks occurs virtually at the same fast mobility $u^2$, 
i.e. the different cage asphericities do not lead to different mobilities.}
 \label{contorPlot_a}
\end{figure}

\begin{figure}[t]
\begin{center}
\includegraphics[width=0.8\linewidth]{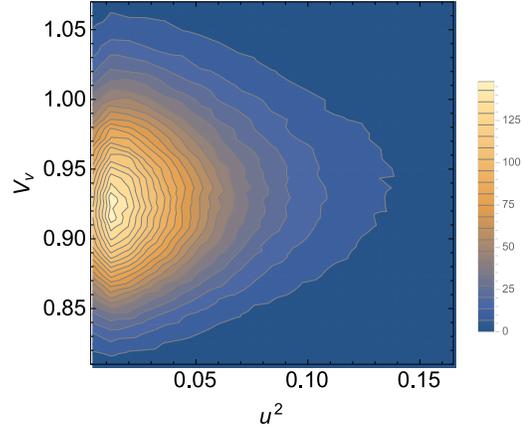} \\
\end{center}
\caption{Joint distribution of the VP volume  $V_v$ and the fast mobility  
$u^2$,  $P(V_v,u^2)$, of the molecular-liquid state state $\color{green} 
\CIRCLE$ (see caption of
Fig.\;\ref{ISF-MSD}).  The two small peaks at $u^2 \simeq  0.012$ with $V_v   
\simeq 0.915$ and $V_v  \simeq 0.93$ are the peaks of the distributions of the 
central and the end monomers of each chain, respectively 
\cite{BarbieriEtAl2004}. Note that the peaks occurs at the same fast mobility 
$u^2$, i.e. the different volumes do not result in different mobilities.}
 \label{contorPlot_v}
\end{figure}

\section{Results}
\label{resultsdiscussion}
\subsection{Fast mobility and relaxation}
\label{dynamic}

\begin{figure}[t]
\begin{center}
\includegraphics[width=0.9\linewidth]{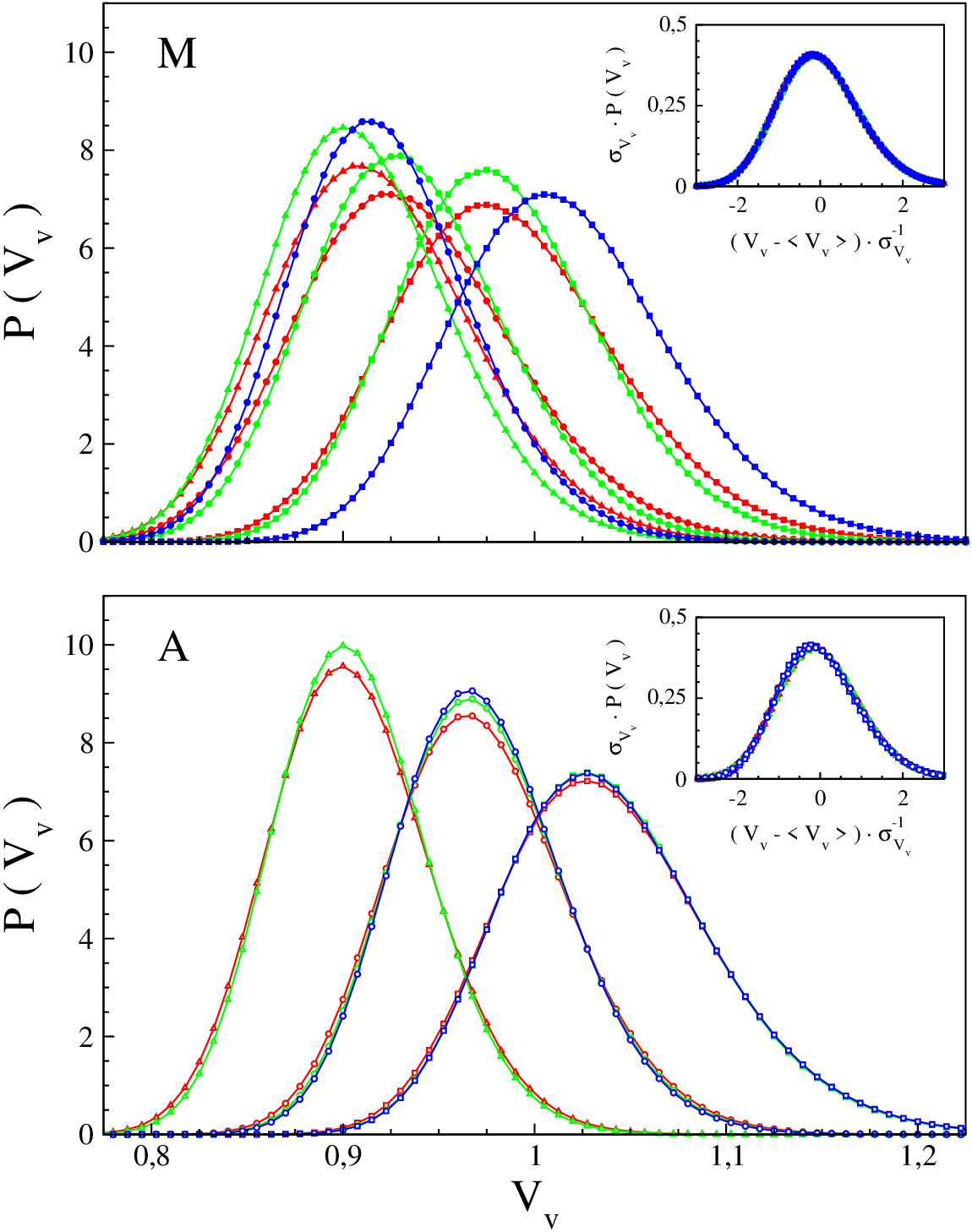} \\
\end{center}
\caption{Distribution of the VP volume $V_v$ of the iso-relaxation time states  
of the molecular (top) and the atomic (bottom) liquids considered in 
Fig.\ref{ISF-MSD} (same symbols and color codes). It is seen that states with 
{\it equal} relaxation time have {\it different} distributions. The inserts 
show that the distributions collapse to a single master curve by proper shift 
and scaling in terms of their average and standard deviation, respectively.}
\label{vol-asf}
\end{figure}

\begin{figure}[t]
\begin{center}
\includegraphics[width=0.9\linewidth]{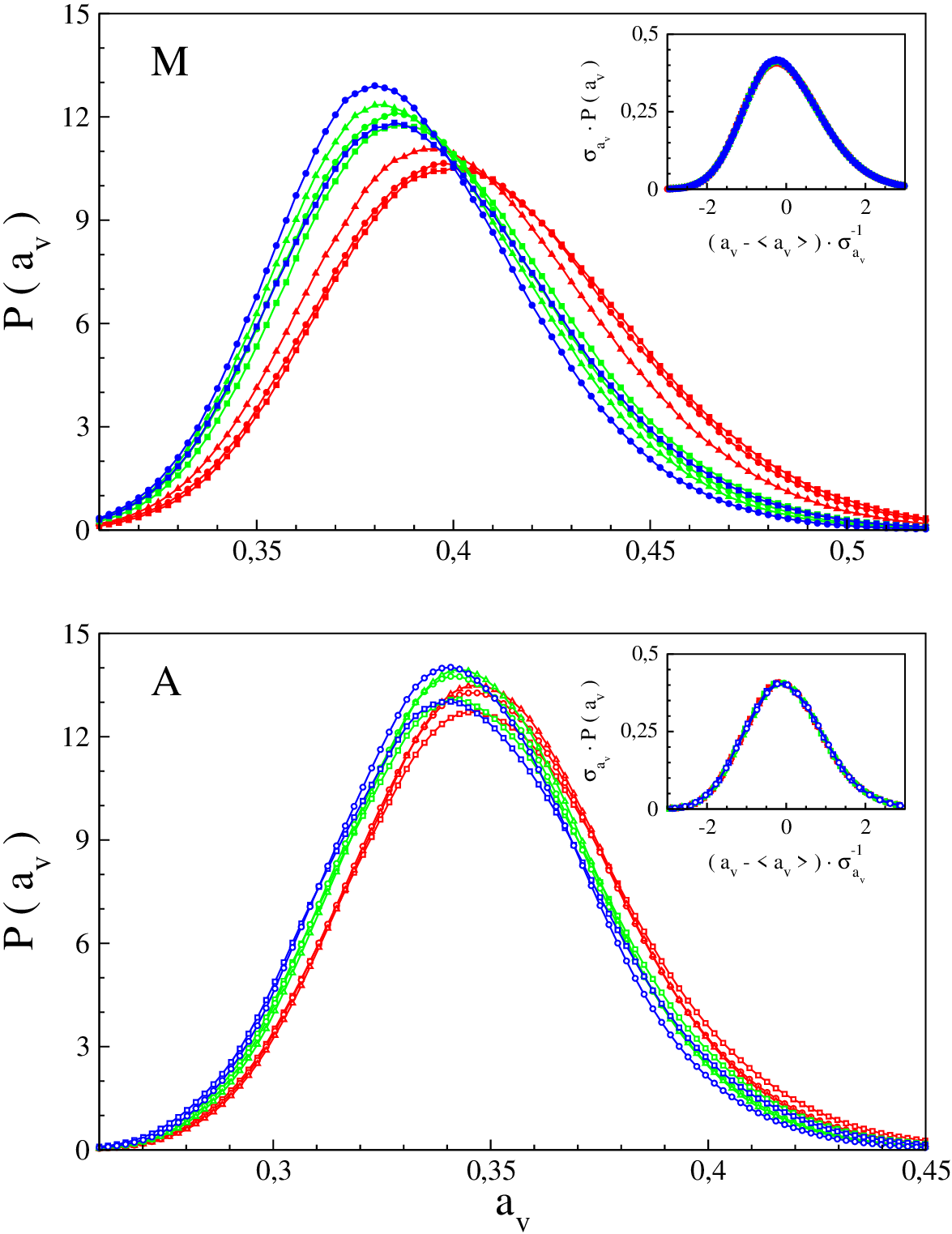} \\
\end{center}
\caption{
Distribution of the VP asphericity $a_v$ of the iso-relaxation time states of  
the molecular (top) and the atomic (bottom) liquids considered in 
Fig.\ref{ISF-MSD} (same symbols and color codes). It is seen that states with 
{\it equal} relaxation time have {\it different} distributions. The inserts 
show that the distributions collapse to a single master curve by proper shift 
and scaling in terms of their average and standard deviation, respectively.}
\label{vol-asf2}
\end{figure}

\noindent
We consider the mean square displacement, i.e. the second moment of the  
incoherent van Hove function:
\begin{equation} \label{MSD}
\langle r^2 (t) \rangle = 4\pi\int_0^\infty r^2 G_s(r,t) \, r^2 dr
\end{equation}
At short times it is known that the derivative $\varDelta (t) = \partial \log  
\langle r^2(t) \rangle / \partial \log t $ exhibits a minimum at $t=t^*$ 
\cite{OurNatPhys,SpecialIssueJCP13,lepoJCP09}. $t^*$ is a measure of the
trapping time of the particle and, in actual units, corresponds to a few 
picoseconds.  In the present molecular liquid model, $t^*_M\cong
1$ and it is virtually independent of the physical state 
\cite{OurNatPhys,lepoJCP09}.  Instead, in the binary mixture $t^*_A$ and 
$t^*_B$ increase slightly with the relaxation time 
\cite{lepoJCP09,SpecialIssueJCP13}. 
We define the fast mobility of the monomers of the linear chains ($i=M$) and 
the  A particles of the atomic mixture ($i=A$) as the short-time mean square 
displacement:
\begin{equation}
 \langle u^2_i \rangle = \langle r^2(t=t^*_i)\rangle, \hspace{5mm} i \in \{A,M \}
 \label{ST-MSD}
\end{equation}
The fast mobility is the mean square amplitude of the position fluctuations of  
the tagged particle in the cage of the neighbours. 
The B particles are not considered. Henceforth,  the $i$ subscript is removed 
from   $ \langle u^2_i \rangle$ for clarity reasons, leaving the identification 
to the context. 

With the purpose of characterizing the long-time structural relaxation we 
consider  the incoherent, self-part of
the intermediate scattering function \cite{HansenMcDonaldIIIEd}:
\begin{eqnarray}
F_{s}(q,t) &=& \int G_{s}(\bm r,t) \exp ( - i \bm q \cdot \bm r) \; d \bm r   
\label{vh-ISF}\\
&=&  4 \pi \int_0^\infty G_{s}(r,t) \, \frac{\sin q r}{q r} \, r^2 d r   
\label{vh-ISF2}
\end{eqnarray}
Eq.\ref{vh-ISF2} follows from the isotropy of the liquids.  The intermediate  
scattering function  provides a convenient function to study the rearrangements 
of the spatial structure of the fluid
over the length scale of $2\pi/q$ . $F_s(q,t)$ is evaluated at the maximum of  
the static structure factor, $q=q^{max}$, with $7.02\leq q^{max}_M \leq 7.39$ 
for the molecular liquid and $7.05\leq q^{max}_A \leq 7.38$ for the A particles 
of the atomic mixture. We are interested in the structural relaxation of the 
molecular liquid ($i=M$) and the A fraction of the binary mixture ($i=A$) and 
define the
structural relaxation time by the equation: 
\begin{equation}  \label{deftau}
F_s(q^{max}_i,\tau_{\alpha i})=e^{-1} \hspace{4mm}  i \in \{A,M \}
\end{equation}
According to Eq.\ref{MSD}, Eq.\ref{ST-MSD}, Eq.\ref{vh-ISF} and 
Eq.\ref{deftau},   two states X and Y fulfilling Eq. \ref{vanhovescaling} 
exhibit {\it equal} fast mobility and {\it equal} relaxation times, i.e. 
Eq.\ref{vanhovescaling2}. 

Selected groups of iso-relaxation time states will be now characterized. They  
are labelled as $M1$, $M2$, $M3$ and $A1$, $A2$, $A3$ and refer to the  
molecular and the atomic liquids, respectively. Details about the states are 
included in the caption of Fig.\;\ref{ISF-MSD} which  shows their intermediate 
scattering function  (top) and mean square displacement (middle). It is seen 
that, after the ballistic regime, the mean square displacement of each set of 
iso-relaxation time states coincide for $t\gtrsim 0.5$, i.e. their fast 
mobility is equal as stated by Eq.\ref{vanhovescaling2}. The bottom panel of 
Fig.\ref{ISF-MSD} shows that the  iso-relaxation time states agree with the 
scaling between the relaxation and the fast mobility 
\cite{OurNatPhys,lepoJCP09,Puosi11,SpecialIssueJCP13}.{Note that a unique 
master curve encompasses both the atomic and the molecular system. 
Fig.\ref{ISF-MSD}(bottom) illustrates the scaling procedure of the fast 
mobility to extend Eq.\ref{vanhovescaling2} and deal with states of different 
physical systems. The procedure is widely discussed elsewhere  
\cite{OurNatPhys,lepoJCP09,Puosi11,SpecialIssueJCP13}}.

Fig.\;\ref{vanHove} plots the incoherent van Hove function $G_s(r,t)$ at the  
trapping (top) and the relaxation times (bottom). It
evidences the strong version of the scaling between the fast trapping regime 
and  the slow relaxation one, as stated by Eq.\ref{vanhovescaling}, namely it 
shows that the coincidence of the incoherent van Hove function $G_s(r,t)$ at 
$t= \tau_\alpha$ of the  iso-relaxation time states implies the coincidence at 
the trapping time $t=t^*$ as well. The coincidence of $G_s(r,t^*)$ implies the 
coincidence of the distributions of the fast mobilities in that:
\begin{equation}
P(u^2) = 2 \pi \, \sqrt{u^2} \, G_s \left (\sqrt{u^2},t^*\right )
\label{Pu2} 
\end{equation}

Note that: i) $G_s(r,\tau_{\alpha})$ of the set of states $M3$ and $A3$  
exhibits the characteristic bump of  the jump dynamics at $r \sim 1$ (the 
particle diameter) and ii) the coincidence of $G_s(r,\tau_\alpha)$ includes the 
exponential tails due to the dynamic heterogeneity \cite{KobPRL07} (see insets 
of Fig.\;\ref{vanHove}).  This makes it apparent that the fast caged dynamics 
is enough to predict key aspects of the dynamical heterogeneity of the glass 
formers \cite{Richert02}, as also seen by previous studies of the breakdown of 
the Stokes-Einstein law \cite{Puosi12SE} and the non-gaussian character of the 
particle displacements \cite{NGP_Wolynes}.

\subsection{Correlations between cage geometry and fast mobility}
\label{correlations}

To characterize the geometry of the cages of the particles, we perform a  
Voronoi tessellation.  In particular, we are interested in both the  volume 
$V_v$  and  the asphericity $a_v$ of VPs. The asphericity is defined as: 
\begin{equation}
 a_v=\frac{\left(A_v\right)^3}{36\pi \left(V_v\right)^2}-1
 \label{aspher}
\end{equation}
where $A_v$ is the VP surface. The asphericity vanishes for a spherical VP  and 
is positive otherwise.  

Our search of the possible correlation between the cage geometry and the fast  
mobility considers the {\it initial} volume $V_v$ and asphericity $a_v$ of the 
VP surrounding a particle at time $t_0$, and compute the {\it subsequent} 
square displacement of the particle  between $t_0$ and $t_0+t^*$, $u^{2}$. The 
ensemble average of $u^{2}$ coincides with Eq.\ref{ST-MSD}.

\subsubsection{Joint distributions  $P(a_v, u^2)$ and $P(V_v, u^2)$}
\label{static-dynamic}

First, the joint probability distribution $P(X, u^2)$ with $X \in \{a_v, V_v 
\}$  is examined.  The distribution is found to be negligibly dependent on the 
physical state and the kind of system, molecular or atomic liquid.

As an example, Fig.\;\ref{contorPlot_a} shows the probability distribution 
$P(a_v,u^2)$  for one state of the molecular liquid. The average is $\langle 
a_v \rangle \simeq 0.39$ which is close to the one of a dodecahedron ($a_v = 
0.325034$), namely the cage asphericity is small as a consequence of the good 
packing of the particles.  Fig.\;\ref{contorPlot_a} strongly suggests that the 
correlation between the fast mobility $u^2$ of the particle and the cage 
asphericity  is rather poor on a per-particle basis. This is not due to the 
"myopic" spatial resolution of $P(a_v,u^2)$. In fact, there is clear evidence 
of two peaks located at $u^2 \simeq  0.012$ and $a \simeq 0.38$ which are seen 
to correspond (not shown)  to the peaks of the distribution $P(a_v,u^2)$ 
restricted to the inner (higher asphericity) and the end (lower asphericity) 
monomers. Fig.\ref{contorPlot_v} shows  the probability distribution 
$P(V_v,u^2)$ for the same state of the molecular liquid. The pattern is very 
similar to the one of $P(a_v,u^2)$. In particular, the peaks of the 
distributions of the central (smaller volume) and the end (larger volume) 
monomers of each chain are seen \cite{BarbieriEtAl2004}. 

The double-peak structure of both Fig.\;\ref{contorPlot_a} and 
Fig.\ref{contorPlot_v}  shows that the inner and the end monomers of the chain 
molecule have {\it equal} fast mobility in spite of the {\it different} size  
\cite{BarbieriEtAl2004} and shape of their cages (see also Fig.6 of 
ref.\cite{VoronoiBarcellonaJNCS14}). This suggests, in addition to the poor 
correlations seen in Fig.\;\ref{contorPlot_a} and Fig.\ref{contorPlot_v}, that 
the fast mobility is not driven only by the geometry of the cage. This aspect 
will be discussed in the next sections  in more detail.

\begin{figure}[t]
\begin{center}
\includegraphics[width=0.9\linewidth]{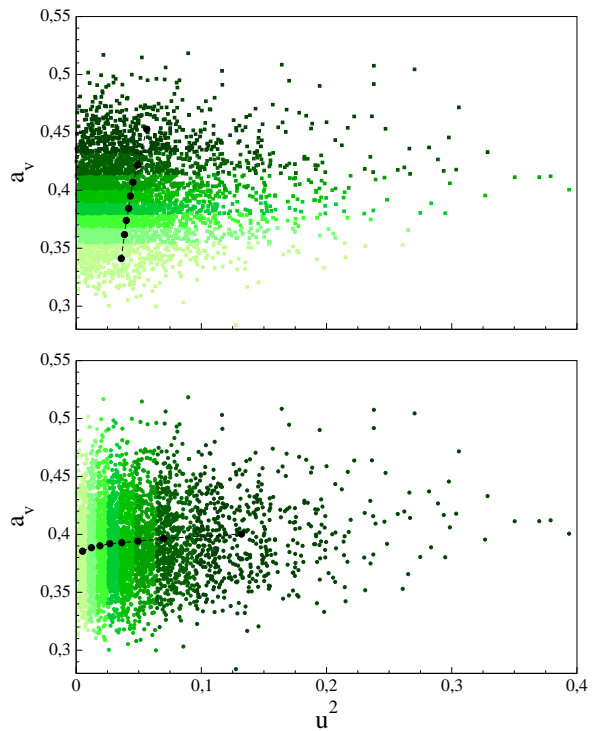} \\
\end{center}
\caption{Subset partition of one ensemble of realisations of the fast 
mobility-asphericity   pairs $\{u^2,a_v\}$  of the molecular-liquid state 
$\color{green} \CIRCLE$ (see caption of Fig.\;\ref{ISF-MSD}). Top: asphericity 
subsets; the black dots have coordinates $\{\langle u^2 \rangle_{a_v,s}, 
\langle a_v \rangle_{a_v,s}\}$ with $s=1,\cdots,8$. $\langle u^2 
\rangle_{a_v,s}$ and $\langle a_v \rangle_{a_v,s}$ are the averages of the fast 
mobility and the asphericity restricted to each set, i.e. Eq.\ref{u2s} and 
Eq.\ref{xs} with  $X = Z = a_v$, respectively. Bottom: fast-mobility subsets; 
the black dots have coordinates $\{\langle u^2 \rangle_{u^2, s}, \langle a_v 
\rangle_{u^2, s}\}$ with $s=1,\cdots,8$. $\langle u^2 \rangle_{u^2, s}$ and 
$\langle a_v \rangle_{u^2, s}$ are the averages of the fast mobility and 
asphericity restricted to each set, i.e. Eq.\ref{u2s} and Eq.\ref{xs} with  $X 
= a_v$, $Z = u^2$, respectively. Note that the average mobility increases with 
the average asphericity in both kinds of partitions.}
\label{Cut-asf}
\end{figure}

\subsubsection{Marginal distributions  $P(a_v)$ and $P(V_v )$ of iso-relaxation states}
\label{static}

\begin{figure}[t]
\begin{center}
\includegraphics[width=0.9\linewidth]{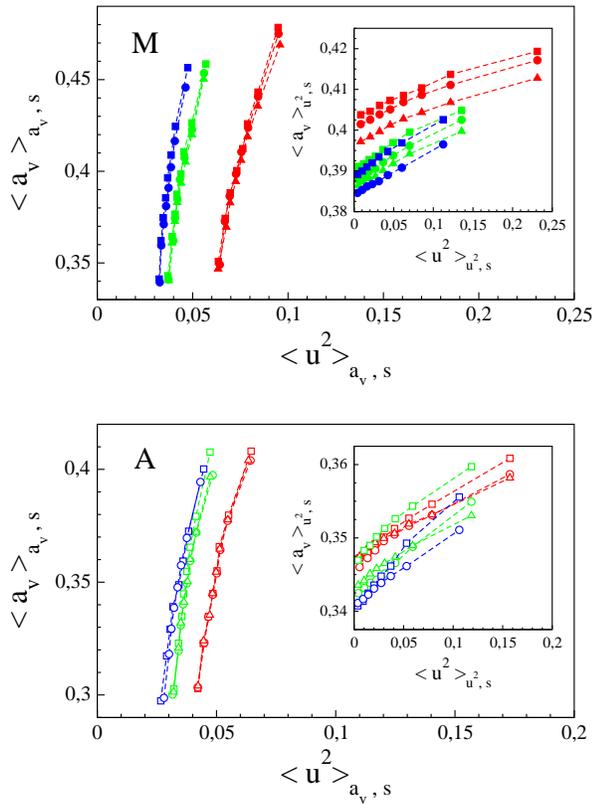} \\
\end{center}
\caption{Relation between the initial average asphericity and the subsequent 
average  fast mobility of the particles belonging to the asphericity subsets 
(see Fig. \ref{Cut-asf} top). Particles are  the monomers of the molecular 
liquid (top) and the A particles of the atomic mixture (bottom).  The inserts 
plot the same relation for the fast-mobility subsets (see Fig. \ref{Cut-asf} 
bottom). In both kind of subsets the mobility increases with the cage 
asphericity in a state-dependent way. Symbols and color codes as in 
Fig.\ref{ISF-MSD}.}
 \label{u2-asfericita}
\end{figure}
\begin{figure}[t]
\begin{center}
\includegraphics[width=0.9\linewidth]{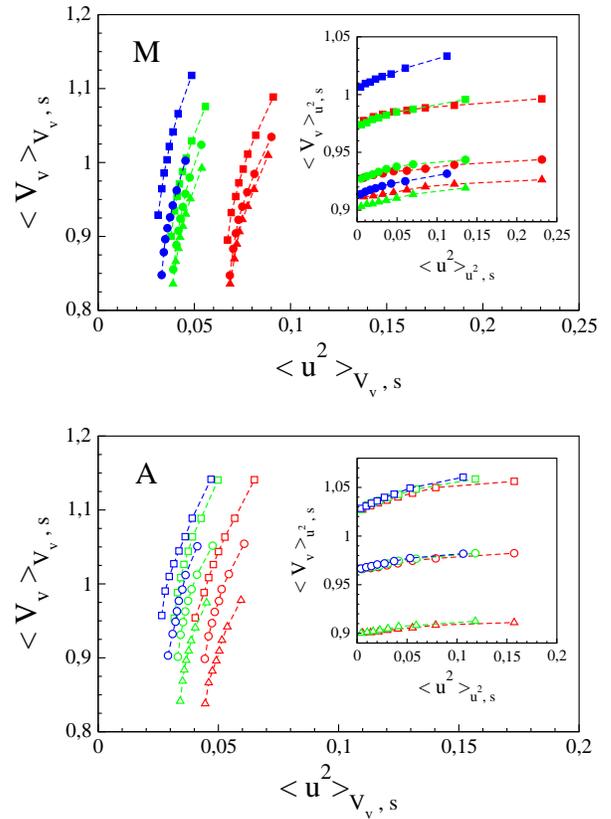} \\
\end{center}
\caption{Relation between the initial average cage volume and the subsequent 
average  fast mobility of the particles belonging to the volume subsets. 
Particles are  the monomers of the molecular liquid (top) and the A particles 
of the atomic mixture (bottom).  The inserts plot the same relation for the 
fast-mobility subsets. In both kind of subsets the mobility increases with the 
cage asphericity in a state-dependent way. Symbols and color codes as in 
Fig.\ref{ISF-MSD}.}
 \label{u2-volume}
\end{figure}

The previous section shows that the joint distribution of the {\it initial} 
local  geometry and the {\it subsequent} fast dynamics of a single state does 
not reveal clear correlations. This suggests the limited insight provided by 
the per-particle analysis. To substantiate the matter, we compare the states 
with equal relaxation time, and then equal average global  fast mobility 
according to Eq.\ref{vanhovescaling2}, which are shown in Fig.\ref{ISF-MSD} and 
Fig. \ref{vanHove}. Clear indications are drawn by examining the marginal 
distributions of the VP volume and asphericity, $P(V_v)$ and  $P(a_v)$ 
respectively:
\begin{eqnarray}
P(X) &=& \int P(X, u^2) \, d u^2, \hspace{5mm} X \in \{ V_v, a_v \}
\end{eqnarray}

Fig.\;\ref{vol-asf} plots the distributions of the VP volume of the monomers of 
 the molecular liquid  (top) and the A particles of the binary mixtures 
(bottom). It is seen that states with {\it equal} relaxation time have {\it 
different} distributions. The differences are traced back to differences in the 
average VP volume ($1/\rho$) and variance. If they are removed by proper 
scaling, the shape of the distributions coincide (see inserts),{thus confirming 
previous claims about its universality \cite{StarrEtAl02}}.  
Fig.\;\ref{vol-asf} suggests that the VP volume is a poor predictor of the 
long-time relaxation. The weak correlations between the VP volume and the 
dynamics has been already noted in polymers \cite{StarrEtAl02}. 

Fig.\;\ref{vol-asf2} plots the distributions of the VP asphericity of the 
monomers  of the molecular liquid  (top) and the A particles of the binary 
mixtures (bottom). Some similarities with the volume distributions are 
apparent, namely: i) the distributions depend on the state and are {\it 
different} even if the states have {\it equal} relaxation time, ii) the 
different distributions collapse on a master curve by proper scaling in terms 
of the average and the variance (see inserts),{thus confirming previous claims 
about its universality \cite{StarrEtAl02}}.  The asphericity distribution 
depends on the state but the changes are much more limited than the ones of the 
VP volume, especially as far as the atomic mixture is concerned, due to the 
good packing. The limited asphericity changes must be contrasted with the wide 
changes of the structural relaxation time (Fig.\ref{ISF-MSD}). We point out 
that the average asphericity is higher for the molecular liquid than the atomic 
mixture. This is ascribed to the bonds of the molecule. In fact, the cage 
surrounding a monomer  is less spherical due to the presence of one or two 
other bonded monomers which are closer than the non-bonded monomers (see 
Sec.\;\ref{numerical})\cite{LocalOrderJCP13}.

\subsubsection{Subset distributions of iso-relaxation states}
\label{static-dynamic2}
According to Sec.\ref{static-dynamic} and Sec.\ref{static}, looking at the 
systems on  particle-by-particle basis does not identify links between the 
rattling of a particle confined by the first neighbours and the cage geometry.

We now show that more insight is drawn by grouping fast mobility and local 
geometries in  subsets.
To this aim, we partition the distributions $P(X, u^2)$ with $X\in \{a_v,V_v 
\}$  in  subsets taken as octiles.  We consider two different kinds of subsets: 
stripes parallel to the $u^2$ axis, resulting in volume or asphericity subsets 
and stripes parallel to the X axis resulting in fast-mobility subsets. 
Fig.\ref{Cut-asf} provides an illustrative example of the two kinds of 
partitions with $X = a_v$.
In general, the $s$-th stripe $[Z_s,Z_{s+1}]$, $Z\in \{a_v,V_v, u^2 \}$, is defined as:
\begin{equation}
\int_{[Z_s,Z_{s+1}]} P(X, u^2) \, dX \, d u^2 = \frac{1}{8}, \, \hspace{2mm}  s=1,\cdots,8
\end{equation}
with $Z_1= 0, \cdots, Z_9 = \infty$. The relevant averages restricted to the $s$-th Z stripe are defined as ($X\in \{a_v,V_v \}$) :
\begin{eqnarray}
\langle u^2 \rangle_{Z_s}  &=& 8 \int_{[Z_s,Z_{s+1}]} u^2 \, P(X, u^2) \, dX \, d u^2  \label{u2s}\\
\langle X \rangle_{Z_s}  &=& 8 \int_{[Z_s,Z_{s+1}]} X \, P(X, u^2) \, dX \, d u^2  \label{xs}
\end{eqnarray}
Both $\langle u^2 \rangle_{Z_s} $ and $\langle X \rangle_{Z_s}$ increase $s=1,\cdots,8$.
As an example, Fig.\ref{Cut-asf} shows the location of the points $\{\langle u^2 \rangle_{a_v,s}, \langle a_v \rangle_{a_v,s}\}$  (top) and  $\{\langle u^2 \rangle_{u^2, s}, \langle a_v \rangle_{u^2, s}\}$ (bottom).
By definition, the relation between the global and the subset averages is:
\begin{equation}
\langle \Re \rangle = \frac{1}{8}\sum_{s=1}^8 \langle \Re \rangle_{Z_s}
\label{avgstripe}
\end{equation}

{Similar divisions in subsets were carried out in previous studies too. The 
partition of  local  structures in subsets according to the VP asphericity has 
been considered for characterising the water structure 
\cite{ShihAsphericityWater} , whereas the correlations between the VP volume 
and the displacements of the fastest and the slowest particles were 
investigated in colloids and polymers  \cite{StarrWeitz05}.}

First,  we investigate all the states in terms of the asphericity and the 
fast-mobility  subsets (for illustration see Fig. \ref{Cut-asf}). 
Fig.\;\ref{u2-asfericita}  plots the {\it initial} average asphericity of the 
cage versus the {\it subsequent} average fast mobility of the two classes of 
subsets.  It is seen that  the mobility increases with the cage asphericity. 
There is no clear signature of this intuitive result in Fig.\ref{contorPlot_a}. 
This is due to the weak correlations which manifest in 
Fig.\;\ref{u2-asfericita}. In fact, $\langle u^2 \rangle_{a_v,s}$ is weakly 
dependent on $\langle a_v \rangle_{a_v,s}$ (Fig.\;\ref{u2-asfericita}, main 
panels), and  
$\langle a_v \rangle_{u^2, s}$ is weakly dependent on $\langle u^2 \rangle_{u^2, s}$ (Fig.\;\ref{u2-asfericita}, inserts).

The subset analysis is now applied to the cage size.  Fig.\;\ref{u2-volume} 
summarizes the results concerning the volume and the fast-mobility subsets as a 
plot of the {\it initial} average volume of the cage versus the {\it 
subsequent} average fast mobility. 
It is seen that  the mobility increases with the cage size in both kind of 
subsets.  Similarly to the asphericity,
there is no clear signature of this second intuitive result in 
Fig.\ref{contorPlot_v}  due to the weak correlations. The weak increase of the 
mobility with the VP volume has been also noted in colloids and polymers  by 
considering 
the displacements of 10 \% fastest and  10 \% slowest particles
\cite{StarrWeitz05}. All in all, Fig.\;\ref{u2-volume} confirms that the VP 
volume is  of modest interest  \cite{StarrEtAl02,HarrowellFreeVolume06}.

One disappointing conclusion drawn by both Fig.\ref{u2-asfericita} and 
Fig.\ref{u2-volume}  is that the increase of the mobility with either the cage 
size or asphericity is {\it state dependent}. In particular, the increase is 
different even if the states have equal relaxation time. Such states have 
identical distribution of the fast mobility, Eq.\ref{Pu2}. Then, if X and Y are 
two iso-relaxation states, one has from Eq.\ref{u2s}:
\begin{equation}
\langle u^2 \rangle_{u^2, s}^{(X)}  = \langle u^2 \rangle_{u^2, s}^{(Y)} \hspace{1mm},   \hspace{3mm} s=1,\cdots,8
\label{u2Z} 
\end{equation}
Eq.\ref{u2Z} states that, if the states have equal relaxation time, the 
fast-mobility  subsets have {\it equal} average value of the fast mobility. 
Nonetheless, the inserts of Fig.\ref{u2-asfericita} and Fig.\ref{u2-volume} 
show that the {\it initial}  average size and shape of the cages embedding 
these particles are {\it different}. 

As a final remark, we remind that both the fast mobility $\langle u^2 \rangle$ 
and the  structural relaxation time have relation with the voids between 
particles, as measured by the Positron Annihilation Lifetime Spectroscopy 
(PALS) \cite{SolesPRL01}. Recently, it was shown that  $\langle u^2 \rangle 
\propto \tau_3$ where $\tau_3$ is the PALS lifetime, which is related to the 
average size of the cavity where the annihilation takes place 
\cite{UnivPhilMag11,OttochianLepoJNCS11}.
Our findings about the poor correlation between the fast mobility and the cage 
size are  seemingly at variance with this experimental results.  However, it 
must be noted that PALS is sensitive to the {\it unoccupied volume} which is 
more closely represented by the volume of the tetrahedra following the Delaunay 
tessellation  \cite{SuterDelaunay92}. Our study is concerned with the Voronoi 
tessellation. The difference between the two tessellations is that the Delaunay 
tetrahedron describes the shape of the cavity between the particles whereas the 
Voronoi polyhedron describes the coordination of the nearest-particles.  The 
investigation of the possible correlation between the fast mobility and the 
size/ shape of the Delaunay tetrahedra is beyond the purpose of the present 
paper.

\subsubsection{Fast mobility and local forces}
\label{forces}
Ultimately, the mobility of a tagged particle is set, in addition to kinetic 
aspects,  by the total force $\bf F$ exerted by the surroundings which, in 
turn, depends on the geometry of the particles arrangement {\it and} the 
interaction potential. Our present results  point to the conclusion that the 
{\it local} geometry correlates poorly with the fast mobility. In order to 
understand if, in spite of the minor role of the cage geometry, states with 
identical fast mobility exhibit common aspects of the total force acting on a 
particle, we investigated the distribution of the modulus $P(F)$ 
\cite{DyreAttractRepulsJPCM13}. In particular, we are 
interested in the mean square force (MSF) $\langle F^2 \rangle$ which includes 
both binary and triplet force contributions \cite{Boon}. MSF is
related to the second frequency  moment  of the velocity correlation function $\Omega_0^2$ :
\begin{equation}
\langle F^2 \rangle  = 3 m \, k_B \,T \, \Omega_0^2
\label{FDT0}
\end{equation}
$m$ being the particle mass \cite{Boon}. $\Omega_0$ is largely contributed by 
local  forces, i.e. the ones between the tagged particle and the 
first-neighbour shell, \cite{Boon} and represents the frequency at which the 
tagged particle would vibrate if it were undergoing small oscillations in the 
potential well produced by the surrounding particles when kept fixed at their 
mean equilibrium positions around the tagged particle 
\cite{HansenMcDonaldIIIEd}. Then, the ratio $\langle F^2 \rangle \, \sigma^2 / 
k_B^2 T^2  = \langle F^2 \rangle / T^2$ is a measure of the local stiffness.

\begin{figure}[t]
\begin{center}
\includegraphics[width=0.9\linewidth]{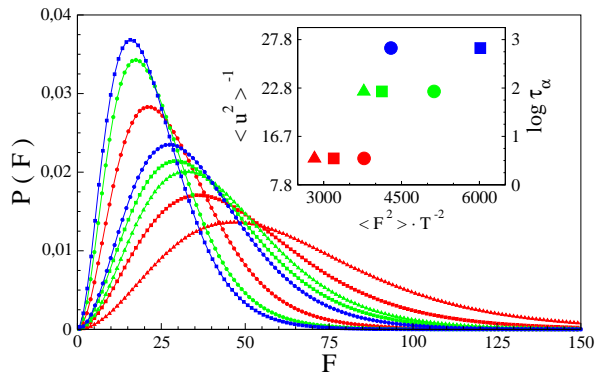} \\
\end{center}
\caption{Distribution of the total force acting on a particle of the molecular  
liquid in the three different sets of  states of Fig.\ref{ISF-MSD} with equal 
fast mobility. The inset shows the poor correlation between the fast mobility 
of the same states with the local stiffness.}
 \label{ForceDistribut}
\end{figure}

Fig. \ref{ForceDistribut} shows that states of the molecular liquid with equal 
fast  mobility have quite different distributions of the total force. They also 
exhibit different local stiffness (Fig. \ref{ForceDistribut}, 
inset). Note also that states with rather different relaxation times have 
rather similar  local stiffness. Even if other states, in particular much 
closer to the glass transition, should be investigated to provide wider 
evidence, we preliminarily conclude that the fast mobility in a 
mildly-supercooled {\it molecular} liquid is not driven by the force exerted by 
the closest neighbours. Notice that iso-mobility and  iso-relaxation time 
states of an {\it atomic} liquid,  with equal density and temperature, exhibit 
identical local force distributions $P(F)$, and then equal MSF 
\cite{DyreAttractRepulsJPCM13}. This different behaviour needs further 
investigation.

\subsubsection{Why poor correlations ?}
\label{why}
The present study reaches three major conclusions: 

\begin{itemize}
\item  there are weak correlations between fast mobility, i.e. the rattling of 
the  particle trapped by the cage of the neighbours, and the cage geometry 
itself. 

\item  the cage geometry is unable to provide a microscopic interpretation of 
the  correlation between the cage rattling and the slow structural relaxation 
as expressed by Eq.\ref{vanhovescaling2}, Eq.\ref{vanhovescaling}. In fact,  it 
is found that two iso-relaxation states have quite clear correlations between 
their fast dynamics, but no obvious relations between their cage geometries. 

\item local forces between a tagged particle and the first-neighbour shell do 
not  correlate with the fast mobility in mildly supercooled states of the 
molecular liquid.

\end{itemize}

Why these failures ? Our results must not lead to the conclusion that no 
correlation  between the structure and the dynamics exists, in that the 
coupling between the dynamic and the structure is reported in the atomic 
mixture \cite{BerthierJackPRE07}  and the molecular liquid \cite{sim} under 
study, also due to the harsh short-ranged repulsive character of the 
potentials.
Instead, we notice that the VP volume and asphericity are mainly controlled by 
the  closest neighbours of the particle and then poorly sensitive to the liquid 
structure beyond the first shell. This aspect matters. The influence of 
structure on dynamics is much stronger on long length scales than on short ones 
in the atomic mixture \cite{BerthierJackPRE07}.

Furthermore, there are evidences in the molecular liquid studied here that the 
fast  motion follows from the collective motion of regions extending more than 
the first coordination shell, a finding which  is consistent 
with elastic models 
\cite{Dyre06,Puosi12,DyreWangGpJCP12,Elastico2,WyartPNAS2013,
DouglasStarrPNAS2015,SchweizerElastic1JCP14,SchweizerElastic2JCP14,
WyartJStatMech07,Harrowell06,Harrowell_NP08,HarrowellJCP09}. One piece of 
evidence comes from the identical fast mobility of {\it separate} parts of the 
molecule, i.e. the end and the central monomers of the molecule, irrespective 
of the different asphericity and size of their cages, see 
Fig.\;\ref{contorPlot_a} and Fig.\ref{contorPlot_v} of the present paper, as 
well as Fig.6 of ref.\cite{VoronoiBarcellonaJNCS14} where the same result is 
found in much longer chain molecules.
Furthermore, it has been evidenced that a given monomer and the surrounding 
particles  up to the next-next-nearest neighbours undergo {\it simultaneous, 
correlated}, rapid displacements within the trapping time $t^*$ over which the 
fast mobility is evaluated (see Eq.\ref{ST-MSD})  
\cite{PuosiLepoJCPCor12,PuosiLepoJCPCor12_Erratum}. {One appealing feature of 
the correlations between the fast displacements is that both their {\it 
strength} and {\it spatial extension} are {\it identical} in states with {\it 
equal} relaxation time {and increase with the latter}, thus suggesting  
non-local interpretations of the link between the cage rattling and the slow 
structural relaxation (Eq.\ref{vanhovescaling2}, Eq.\ref{vanhovescaling}) 
\cite{PuosiLepoJCPCor12,PuosiLepoJCPCor12_Erratum}}.
The non-local  view is  in harmony with the finding that the 
structural  relaxation does not correlate with the infinite-frequency elastic 
response, which is {\it local} in nature, but with the {\it quasi-static} 
elastic response at zero wavevector (k=0) which occurs under {\it mechanically 
equilibrium} (${\bf F}=0$) \cite{Puosi12,DyreWangGpJCP12}. This has important 
implications. In fact, the deformation field at low frequency is quite extended 
and, indeed, is scale-free in simple examples , e.g. it decays as $r^{-x}$ with 
$x = 1$ for a localised force and $x = 2$ for cavity expansion. 

The suggested presence of non-local fast collective motion agrees with the 
evidence  of quasi-local soft modes, i.e. with an extended component, in MD 
simulations of  2D binary mixtures and the finding that the regions of motion 
of the quasi-local soft modes exhibit striking correlation with the regions of 
high Debye-Waller factor  \cite{Harrowell06,Harrowell_NP08,HarrowellJCP09}. 
{Moreover, it has been proposed that structural relaxation in deeply 
supercooled liquids proceeds via the accumulation of Eshelby events, i.e. local 
rearrangements that create long-ranged and anisotropic stresses in the 
surrounding medium \cite{LemaitrePRL14}}.

\section{Conclusions}
\label{conclusions}
The present MD study aims at clarifying the role of the cage geometry in 
affecting  the fast mobility $\langle u^2 \rangle$, namely the rattling 
amplitude of the particles temporarily trapped by the cage of the neighbors,  
in mildly supercooled states.  The fast mobility is a very 
accurate predictor of the long-time relaxation, as summarized  by 
Eq.\ref{vanhovescaling2}, stating that two states with equal fast mobility have 
equal relaxation time as well 
\cite{OurNatPhys,lepoJCP09,Puosi11,SpecialIssueJCP13,UnivSoftMatter11,
DouglasCiceroneSoftMatter12,UnivPhilMag11,CommentSoftMatter13,
OttochianLepoJNCS11,Puosi12SE,PuosiLepoJCPCor12,PuosiLepoJCPCor12_Erratum}, and 
the more general Eq.\ref{vanhovescaling} linking the single-particle fast 
dynamics and slow relaxation \cite{Puosi11,SpecialIssueJCP13}. The cage 
geometry has been characterised by the volume $V_v$ and the asphericity $a_v$ 
of the Voronoi polyhedra following the tessellation of the configurations  of 
both a molecular and an atomic liquid.{ A novel cross-correlation analysis 
between  the fast mobility and the Voronoi volume and shape has been 
performed.}

Signatures of connection between the fast mobility and the cage geometry are 
not  found in either the joint ($P(a_v, u^2)$ and $P(V_v, u^2)$) or the 
marginal ($P(a_v)$ and $P(V_v)$) distributions. 
Weak correlations are identified by partitioning the Voronoi cells in 
asphericity  and size subsets, as well as by dividing the particles in 
fast-mobility subsets. The procedure reveals the increase of the average 
fast-mobility with the average size and asphericity of the different subsets. 
The observed correlations are not the same in states with equal relaxation 
time, leading us to the conclusion that the cage geometry is unable to provide 
a microscopic interpretation of the link between the cage rattling and the slow 
structural relaxation, as expressed by Eq.\ref{vanhovescaling2}, 
Eq.\ref{vanhovescaling}. Local forces between a tagged particle and the 
first-neighbour shell do not correlate  with the fast mobility in the molecular 
liquid. We suggest that the single-particle fast mobility follows by modes 
extending beyond  the first neighbours. 

\begin{acknowledgments}
A generous grant of computing time from IT Center, University of Pisa and  M. 
Righini, ${}^\circledR$Intel is gratefully acknowledged.
\end{acknowledgments}


\begin{thebibliography}{113}
\bibitem[{not()}]{note3}
\bibinfo{howpublished}{Note that $t^\star$ is one order of magnitude shorter
  than the times considered in ref. \cite{Harrowell06}.}
\end{thebibliography}
\end{document}